\documentstyle[prd,aps,epsfig]{revtex}
\def\beq{\begin{equation}}
\def\eeq{\end{equation}}
\def\beeq{\begin{eqnarray}}
\def\eeeq{\end{eqnarray}}
\def\oh{\omega}
\def\oq{\bar\omega}
\def\bla{\bar\Lambda}
\def\bl2{\frac{\bar\Lambda}2}

\def\eps{\varepsilon}
\begin{document}
\title{Heavy quark expansion and universal form factors in quark model}
\author{ Dmitri Melikhov}
\address{Nuclear Physics Institute, Moscow State University, 
Moscow, 119899, Russia}
\maketitle

\begin{abstract}
Meson transition amplitudes of the vector, axial--vector, and tensor quark currents are analyzed within 
dispersion formulation of the constituent quark model. 
The form factors in the decay region are given by relativistic double spectral
representations through the wave functions of the inital and final mesons. 
We perform heavy quark expansion of the quark--model mesonic matrix elements  
with a next--to--leading order accuracy and 
demonstrate that matching this expansion to the heavy--quark expansion in QCD requires subtractions in 
the double spectral representations for the form factors and 
allows fixing the subtraction terms. The Isgur--Wise function 
and next--to--leading order universal form factors are calculated. 
\end{abstract}

\section{Introduction}
Theoretical description of hadronic amplitudes of the quark currents is one of the key problems of particle 
physics as such amplitudes provide a bridge between QCD formulated in the language of quarks and gluons and
observable phenomena which deal with hadrons. In particular, the knowledge of such amplitudes is
necessary for extraction of the parameters of the quark--mixing matrix in the Standard Model from the
experiments on weak heavy hadron decays. The difficulty in such calculations lie in the fact that hadron
formation occurs at large distances where perturbative QCD methods are not applicable and nonperturbative
consideration is necessary.

Various theoretical frameworks more or less directly related to QCD have been applied to the description
of meson transition form factors: among them constituent quark models 
\cite{wsb,altomari,koerner,isgw,isgw2,jaus,stech,faustov,beyer,pene,grach,cheng,demchuk,m1,m2}, 
lattice QCD simulations \cite{lat1,lat2,lat3}, 
QCD sum rules \cite{bbd,ball,narison,colangelo,lcsr1,lcsr2,lcsr3}, 
analytical constraints \cite{anal1,anal2}. 

Lattice QCD simulations is the most direct QCD based nonperturbative approach and thus should in principle 
provide most reliable results. 
However it faces serious technical problems with placing heavy particles on the lattice and performing 
calculations with inclusion of quark loops. So far direct calculations with $b$-quark are not possible and 
extrapolation in the quark mass is used which considerably reduces the predictive power of the method. 

Various versions of QCD sum rules to meson decays give very uncertain predictions strongly dependent 
on the technical subtleties of the particular version. A recent analysis \cite{lcsr1} disregards the 
three--point QCD sum rules in favor of the light--cone sun rules which however involve more phenomenological 
inputs.

Constituent quark models have proved to be a fruitful phenomenological method for the description of 
heavy meson transition form factors. Constituent quark models employ the fact that heavy meson consists of a
heavy quark and light degrees of freedom with have quantum numbers of the quark state and assume that these 
light states can be approximately described by an effective constituent quark. 
The application of the various versions of the constituent quark picture to the decay processes
has a long history.  
The first models were based on a semirelativistic \cite{wsb} or 
nonrelativistic \cite{altomari,koerner,isgw} considerations and didn't treat the quark spins thoroughly. 
Further developments 
demonstrated the relativistic description of the quark spins to be important for consistent 
applications to meson decays. The exact solution to a complicated dynamical problem 
of treating the spins of the interacting particles is not known, 
but a simplified self--consistent relativistic consideration of the quark spins can be performed within 
the light--cone quark model (LCQM) \cite{jaus}.  
The difficulty with the application of this approach to the decay processes lies in the fact 
that the applicability of the model is 
restricted by the condition $q^2\le 0$, while the physical region for hadron decays is 
$0\le q^2\le (M_i-M_f)^2$, $M_{i,f}$ being the initial and final hadron masses, respectively. 
The problem is connected with the contribution of the so--called pair--creation subprocesses
which cannot be taken into account thoroughly in the model, except for a trivial case of a pointlike 
interaction.  
At spacelike momentum transfers the contribution of these subprocesses can be 
killed by choosing an appropriate reference frame, whereas at timelike momentum transfers 
such a frame does not exist and thus pair creation contributes together with the partonic contribution. 
An estimate of the form factors at timelike momentum transfers within the LCQM has been done in \cite{jaus} 
where the transition form factors have been calculated in the spacelike region and then numerically 
extrapolated to the timelike region assuming some particular $q^2$--behavior. 
This works well if the accessible momentum transfer interval is not large. 
However, in the heavy--to--light meson decays this interval is of order of the heavy meson mass squared and 
the extrapolation procedure yields large uncertainties. 
The approach of refs \cite{grach,cheng,demchuk} calculates the partonic contribution to the form factors at 
timelike momentum transfers and thus avoids extrapolations. However, the nonpartonic contribution is omitted. 
Such treatment can be justified if the latter is small that is not obvious $a\;priori$. 
Thus, one can see that for a reliable description of the decay processes within the LCQM
it might be reasonable to find another formulation of the model appropriate also at timelike momentum transfers. 
Such a formulation has been proposed in \cite{m1}. 

The approach of \cite{m1} is based on the dispersion formulation of the LCQM. 
Namely, the transition form factors obtained within the light--cone formalism at $q^2<0$ 
\cite{jaus}, are represented as dispersion 
integrals over initial and final hadron masses through their light--cone wave functions. 
The spectral densities of these representations can be calculated from the Feynman graphs. 
The transition form factors at $q^2>0$ are derived by performing the analytical continuation in $q^2$ from the 
region $q^2\le 0$. As a result, for a decay caused by the weak transition of the quark $Q(m_i)\to Q(m_f)$, 
form factors in the region $q^2\le (m_i-m_f)^2$ are expressed through the light--cone 
wave functions of the initial and final hadrons and can be calculated in the decay region avoiding dangerous 
extrapolations. 

At the same time, such a dispersion formulation of the LCQM allows also a bit different view on the problem of constructing the
form factors within the quark model: actually, we calculate the double spectral densities from the Feynman 
graphs and thus obtain unsubtracted spectral representations. On the other hand we know that in general subtractions 
might be added to such representations and we need some additional arguments to decide on the necessity and the structure 
of the subtraction terms. 

Let us notice that once we are working within an approach not directly
deduced from QCD it is important to preserve essential features of the underlying fundamental theory 
in the model. Thus matching the results obtained within the quark model to rigorous QCD results might be 
helpful for bringing more realistic features to the model. We use such matching to QCD results for determining the 
subtraction terms in the double spectral representations of the form factors. 

There are few QCD predictions on meson form factors in the decay region. Fortunately, 
there is an important case in which QCD provides rigorous and model--independent results on transition 
form factors: namely, meson decays induced by transitions between heavy quarks. 
In this case the heavy--quark symmetry simplifies all the form factors in the leading $1/m_Q$ order to the
universal Isgur--Wise function \cite{iwhh}. 
A systematic expansion in inverse powers of the heavy quark mass 
can be constructed within the Heavy quark effective theory \cite{neubert} based on QCD with the heavy quarks.
Although the form factors appearing in different $1/m_Q$ orders can not be determined within the HQET, 
the latter restricts the structure of the expansion and the number of these universal form factors in each $1/m_Q$
order and provides constraints on some of them at zero recoil. 

The idea of matching the quark model form factors to 
the structure of the heavy--quark expansion of HQET has been employed in the ISGW2 model \cite{isgw2}. 
However the ISGW2 model does not calculate the form factors within a relativistic dynamical approach on the 
basis of considering the meson structure but rather gives prescription for constructing the form factors. 
The approach of ref. \cite{stech} also employs matching to the heavy--quark expansion and obtains interesting results 
within the constutuent quark picture 
assuming a strong peaking of the momentum distribution of the constituent quarks inside a meson and avoiding 
details of the dynamics of the process. This model imposes constraints on the form factors but does not allow one 
to calculate them. 

The idea implemented in this work is to perform the heavy--quark expansion in the dispersion 
formulation of the quark model and to match this expansion to the HQET. 
This matching is used for determining the structure of the subtraction terms in the spectral
representations. We consider the form factors of pseudoscalar meson transitions into pseudoscalar and vector
final mesons through the vector, axial--vector, and currents and analyze the expansion in the leading and 
next--to--leading order. 
Further constraints on the subtraction terms are obtained by considering the case of the 
heavy--to--light quark transition and analyzing the relations between the form factors of the tensor, vector 
and axial--vector currents. Comparing the relations obtained within the dispersion quark model to the general 
relations found by Isgur and Wise \cite{iwhl} allows us to fix the subtraction terms. 

Our main results are the following: \\
1. Applying the approach of \cite{luke} to heavy meson transition throught the tensor current, we derive the
decomposition of the form factors $h_{g_+}$, $h_{g_-}$, $h_{g_0}$, and $h_{s}$ (the definitions are given in
the next section) in the next--to--leading order in $1/m_Q$. In particular, we have found that the $1/m_Q$
corrections to the form factor $h_{g_+}$ at zero recoil vanish just as for the form factors $h_{f_+}$ and
$h_f$ (the Luke theorem). 
Although the heavy meson transitions through the tensor current do not correspond to a real experimental 
situation we need the expansion for such form factors as a bench mark for comparison of the quark model. \\
2. We present an improved dispersion formulation of the LCQM: the transition form factors of pseudoscalar ($P$)
mesons into pseudoscalar ($P$) and vector ($V$) mesons induced by the vector, axial--vector 
and tensor currents are given by double spectral representations through the wave functions of the 
initial and final mesons. The unsubtracted double spectral densities are calculated from the Feynman graphs 
and the subtraction terms in the spectral representations are determined by matching the $1/m_Q$ expansion in
the quark model to the HQET. \\
3. We perform the heavy--quark expansion of the form factors of the dispersion quark model up to NLO and calculate the 
Isgur--Wise function and the NLO universal form factors assuming a strong peaking of the soft wave function 
near the $q\bar q$ threshold with a width of order $\Lambda_{QCD}$. We obtain in the LCQM the relations 
\beq
\label{nloffs}
\chi_2(\oh)=0;\qquad \chi_3(\oh)=0;\qquad \xi_3(\oh)>0,\; \xi_3(1)={\langle z \rangle}/{3}, 
\eeq
where $\langle z \rangle$ is an average kinetic energy of the light quark in the heavy meson rest frame. 

We observe that matching the $1/m_Q$ expansion of the quark model to HQET in LO and NLO requires
subtractions in some of the form factors describing the $P\to V$ transition and
restricts the form of the subtraction terms.\\
4. We consider the heavy--to--light meson transitions in which case a small parameter 
$\Lambda_{QCD}/m_Q$ emerges and analyze the form factors in the leading $\Lambda_{QCD}/m_Q$ order.
Requiring the fulfillment of the Isgur--Wise relatins for the heavy--to--light transitions \cite{iwhl} 
further constrains the subtraction terms providing explicit spectral representations with subtractions 
for the form factors of interest.

In the next section we remind the definitions of the form factors and list the HQET results for the amplitudes 
of the $P\to P$ and $P\to V$ transitions through the vector, axial--vector and tensor currents up to NLO. 
In section 3 we present the quark--model results for the meson decay form factors at arbitrary masses: 
namely, we calculate the unsubtracted spectral densities and explicit form of the subtraction terms which 
form is motivated in the subsequent section. In Section 4 the $1/m_Q$ expansions of the 
wave functions and the form factors in quark model for the heavy--to--light transitions are considered. 
In Section 5 the heavy--to--light case is discussed. Section 6 illustrates the main results with numerical
estimates and evaluate the Isgur--Wise function for various quark model parameters. Conclusion 
summarizes the results. 
\section{Meson transition amplitudes and heavy--quark expansion in QCD}
The amplitudes of meson decays induced by the quark transition $q_2\to q_1$ through the vector 
$V_{\mu}={\bar q_1}\gamma_{\mu}q_2$, axial--vector 
$A_{\mu}={\bar q_1}\gamma_{\mu}\gamma_{5}q_2$, and tensor $T_{\mu\nu}={\bar q_1}\sigma_{\mu\nu}q_2$ 
currents have the following structure \cite{iwhl}
\begin{eqnarray}
\label{ffdef}
<P(M_2,p_2)|V_\mu(0)|P(M_1,p_1)>&=&f_1(q^2)p_{1\mu}+f_2(q^2)p_{2\mu},  \nonumber \\
<V(M_2,p_2,\epsilon)|V_\mu(0)|P(M_1,p_1)>&=&2g(q^2)\epsilon_{\mu\nu\alpha\beta}
\epsilon^{*\nu}\,p_1^{\alpha}\,p_2^{\beta}, \nonumber \\
<V(M_2,p_2,\epsilon)|A_\mu(0)|P(M_1,p_1)>&=&
i\epsilon^{*\alpha}\,[\,f(q^2)g_{\mu\alpha}+a_1(q^2)p_{1\alpha}p_{1\mu}+
               a_2(q^2)p_{1\alpha}p_{2\mu}\,],   \nonumber \\
<P(M_2,p_2)|T_{\mu\nu}(0)|P(M_1,p_1)>&=&-2i\,s(q^2)\,(p_{1\mu}p_{2\nu}-p_{1\nu}p_{2\mu}), \nonumber \\
<V(M_2,p_2,\epsilon)|T_{\mu\nu}(0)|P(M_1,p_1)>&=&i\epsilon^{*\alpha}\,
               [\,g_{1}(q^2)\epsilon_{\mu\nu\alpha\beta}p^{1\beta}+
                g_{2}(q^2)\epsilon_{\mu\nu\alpha\beta}p^{2\beta}+
                g_0(q^2)p_{1\alpha}\epsilon_{\mu\nu\beta\gamma}p_1^{\beta}p_2^{\gamma}\,], 
\end{eqnarray}
with $q=p_{1}-p_{2}$, $P=p_{1}+p_{2}$. We use the notations: 
$\gamma^{5}=i\gamma^{0}\gamma^{1}\gamma^{2}\gamma^{3}$, 
$\sigma_{\mu \nu}={\frac{i}{2}}[\gamma_{\mu},\gamma_{\nu}]$,
$\epsilon^{0123}=-1$, and 
$Sp(\gamma^{5}\gamma^{\mu}\gamma^{\nu}\gamma^{\alpha}\gamma^{\beta})=
4i\epsilon^{\mu\nu\alpha\beta}$. 

The relativistic--invariant form factors contain the dynamical 
information on the process and should be calculated within a nonperturbative approach 
for any particular initial and final mesons. 

For analysing the transition in the case when both the parent and the daughter quarks 
inducing the meson transition are heavy, i.e. $m_1\simeq m_2\gg \Lambda_{QCD}$  
it is convenient to introduce a new dimensionless variable 
$\oh=v_1v_2=\frac{M_1^2+M_2^2-q^2}{M_1M_2}$ and  
velocity--dependent form factors connected with 4--velocities 
and not 4--momenta as in (\ref{ffdef}) in the following way 
\begin{eqnarray}
\label{hqffdef}
<P(M_2,p_2)|V_\mu(0)|P(M_1,p_1)>&=&\sqrt{M_1M_2}\,[h_{f_+}(\oh)(v_1+v_2)_{\mu}
+h_{f_-}(\oh)(v_1-v_2)_{\mu}],  
\nonumber \\
<V(M_2,p_2,\epsilon)|V_\mu(0)|P(M_1,p_1)>&=&\sqrt{M_1M_2}\;h_g(\oh)\epsilon_{\mu\nu\alpha\beta}
\epsilon^{*\nu}\,v_1^{\alpha}\,v_2^{\beta}, 
\nonumber \\
<V(M_2,p_2,\epsilon)|A_\mu(0)|P(M_1,p_1)>&=&
i\epsilon^{*\alpha}\,\sqrt{M_1M_2}\,
[h_f(\oh)(1+\oh)g_{\mu\alpha}-h_{a_1}(\oh)v_{1\alpha}v_{1\mu}
-h_{a_2}(\oh)v_{1\alpha}v_{2\mu}\,],   \nonumber \\
<P(M_2,p_2)|T_{\mu\nu}(0)|P(M_1,p_1)>&=&-2i\,\sqrt{M_1M_2}\;h_s(\oh)\,(v_{1\mu}v_{2\nu}-v_{1\nu}v_{2\mu}), 
\nonumber \\
<V(M_2,p_2,\epsilon)|T_{\mu\nu}(0)|P(M_1,p_1)>&=&i\epsilon^{*\alpha}\sqrt{M_1M_2}\,
               [\,h_{g_+}(\oh)\epsilon_{\mu\nu\alpha\beta}(v_1+v_2)^{\beta}+
                h_{g_-}(\oh)\epsilon_{\mu\nu\alpha\beta}(v_1-v_2)^{\beta}\nonumber\\
               &&+h_{g_0}(\oh)v_{1\alpha}\epsilon_{\mu\nu\beta\gamma}v_1^{\beta}v_2^{\gamma}\,]. 
\end{eqnarray}
These form factors are related to the form factors introduced by the relations (\ref{ffdef}) as follows
\begin{eqnarray}
\label{ff2hqff}
f_1&=&\frac{M_2}{\sqrt{M_1M_2}}[h_{f_+}+h_{f_-}],\quad
g_1=  \frac{M_2}{\sqrt{M_1M_2}}[h_{g_+}+h_{g_-}],\quad
\nonumber\\
f_2&=&\frac{M_1}{\sqrt{M_1M_2}}[h_{f_+}-h_{f_-}],\quad
g_2=  \frac{M_1}{\sqrt{M_1M_2}}[h_{g_+}-h_{g_-}],\quad
\nonumber\\
g&=&\frac{1}{2\sqrt{M_1M_2}}h_g,\quad  
s=\frac{1}{2\sqrt{M_1M_2}}h_s,\quad  
\nonumber\\
f&=&\sqrt{M_1M_2}(1+\oh)h_f,\quad
a_2=-\frac{1}{\sqrt{M_1M_2}}h_{a_2},\quad 
a_1=-\frac{1}{\sqrt{M_1M_2}}\frac{M_2}{M_1}h_{a_1},\quad 
g_0=\frac{1}{\sqrt{M_1M_2}}\frac{1}{M_1}h_{g_0}
\end{eqnarray}
The form factors $h$ are convenient quantities as in the leading $1/m_Q$ order 
all of them are expressed through a single universal function of the dimensionless 
variable $\oh$ -- the Isgur--Wise function. 
A consistent heavy--quark expansion of the form factors, i.e. expansion in inverse powers of the heavy--quark mass, 
can be constructed within the Heavy Quark Effective Theory based on QCD with heavy quarks. 

The general structure of the $1/m_Q$ expansion of the heavy quark form factors 
in QCD for the meson transition $M_1\to M_2$ induced by heavy quark transition $m_2\to m_1$ have the 
form (omitting corrections $O(\alpha_s,\,\alpha_s/m_Q,\,1/m_Q^2)$:
\begin{eqnarray}
\label{hqexp}
h_{f_+}=&\xi+     &\left(\frac{1}{m_1}+\frac{1}{m_2}\right)\rho_1,\nonumber\\
h_{f_-}=&         &\left(\frac{1}{m_1}-\frac{1}{m_2}\right)\left(-\bl2\xi+\xi_3\right),\nonumber\\
h_{g}=  &\xi+     &\left(\frac{1}{m_1}+\frac{1}{m_2}\right)\bl2\xi+\frac1{m_1}\rho_2
                            +\frac1{m_2}\left({\rho_1-\xi_3}\right),\nonumber\\
h_{f}=  &\xi+     &\left(\frac{1}{m_1}+\frac{1}{m_2}\right)\frac{\oh-1}{\oh+1}\bl2\xi+\frac1{m_1}\rho_2
                            +\frac1{m_2}\left({\rho_1-\frac{\oh-1}{\oh+1}\xi_3}\right),\nonumber\\
h_{a_1}=&         &\frac1{m_1}\frac1{\oh+1}\left(-\bar\Lambda\xi +2(\oh+1)\chi_2-\xi_3\right),\nonumber\\
h_{a_2}=&\xi+     &\left(\frac{\oh-1}{\oh+1}\frac{1}{m_1}+\frac{1}{m_2}\right)\bl2\xi+
\frac1{m_1}\left(\rho_2-2\chi_2-\frac1{\oh+1}\xi_3\right)+\frac1{m_2}(\rho_1-\xi_3).
\nonumber\\
h_{s}=&\xi+       &\left(\frac{1}{m_1}+\frac{1}{m_2}\right)\left({\bl2\xi-\xi_3+\rho_1}\right),\nonumber\\
h_{g_+}=&-\xi-    &\frac1{m_2}\rho_1-\frac1{m_1}\rho_2,\nonumber\\
h_{g_-}=&         &\left(\frac{1}{m_1}-\frac{1}{m_2}\right)\bl2\xi+\frac1{m_2}\xi_3,\nonumber\\
h_{g_0}=&         &\frac1{m_1}\left({    \frac{\bar\Lambda\xi+\xi_3}{\oh+1}+2\chi_2}\right).
\end{eqnarray}
In the leading $1/m_Q$ order (LO) all the form factors are represented through the single universal
Isgur--Wise function $\xi(\oh)$ , whereas in the next-to-leading order (NLO) the 4 new form factors 
$\rho_1, \rho_2, \chi_2$, and $\xi_3$ appear. 
The universal form factors are functions of a single variable $\oh$. 

The form factor $\xi_3$ originates from the expansion of the transition quark current, and the form factors 
$\rho_1,\rho_2,\chi_2$ are connected with the nontrivial relationship between the mesonic states in the full and the
effective theory. The universal form factors satisfy the conditions
\begin{equation}
\label{hqnorm}
\xi(1)=1,\quad \rho_1(1)=\rho_2(1)=0,
\end{equation}
whereas no constraints on $\xi_3$ and $\chi_2$ are imposed by the heavy quark symmetry. 
As found by Luke \cite{luke}, the $1/m_Q$ corrections to the form factors $h_f$ and $h_{f_+}$ vanish due to kinematical or
dynamical reasons. One can see that the same is true for the form factor $h_{g_+}$: namely,
\begin{equation}
h_{g_+}(1)=-1+O(1/m_Q^2).
\end{equation}

The parameter $\bar\Lambda$ in (\ref{hqexp}) comes from the $1/m_Q$ expansion of the mass of a meson consisting of the 
heavy quark and light degrees of freedom
\begin{equation}
\label{lambdadef}
M_Q=m_Q+\bar\Lambda+O(1/m_Q). 
\end{equation}
In our notations for heavy quarks and mesons, this gives 
\begin{equation}
M_1=m_2+\bar\Lambda+\ldots,\qquad M_2=m_1+\bar\Lambda+\ldots,
\end{equation}
for the parent and daughter particles, respectively. 

It is straightforward to derive the following useful relations
\begin{eqnarray}
\label{M2m}
\frac{M_1+M_2}{\sqrt{M_1M_2}}&=&\frac{m_2+m_1}{\sqrt{m_1m_2}}\left[{1-\left(\frac1m_1+\frac1m_2\right)
\left(\frac{m_2-m_1}{m_2+m_1}\right)^2\bl2+\ldots}\right]\nonumber\\
\frac{M_1-M_2}{\sqrt{M_1M_2}}&=&\frac{m_2-m_1}{\sqrt{m_1m_2}}
\left[{1-\left( \frac1{m_1}+\frac1{m_2}\right)\bl2+\ldots }\right]\nonumber\\
\sqrt{M_1M_2}&=&\sqrt{m_1m_2}
\left[{1+\left( \frac1{m_1}+\frac1{m_2}\right)\bl2+\ldots }\right],
\end{eqnarray}
where the dots denote higher order terms. 

Using the relations (\ref{ff2hqff}) and (\ref{M2m}), we obtain for the form factors (\ref{ffdef}) the following
expansions  
\begin{eqnarray}
\label{f1exp}
f_1&=&\frac{m_1}{\sqrt{m_1m_2}}\left[{\xi+\frac1{m_1}\left(\rho_1+\xi_3\right) +\frac1{m_2}\left(\rho_1-\xi_3\right)}\right],\\ 
\label{f2exp}
f_2&=&\frac{m_2}{\sqrt{m_1m_2}}\left[{\xi+\frac1{m_1}\left(\rho_1-\xi_3\right) +\frac1{m_2}\left(\rho_1+\xi_3\right)}\right],\\
\label{sexp}
  s&=&  \frac{1}{2\sqrt{m_1m_2}}\left[{\xi+\frac1{m_1}\left(\rho_1-\xi_3\right) +\frac1{m_2}\left(\rho_1-\xi_3\right)}\right],\\
\label{g1exp}
g_1&=&-\frac{m_1}{\sqrt{m_1m_2}}\left[{\xi+\frac1{m_1}\rho_2+\frac1{m_2}\left(\rho_1-\xi_3\right)}\right],\\ 
\label{g2exp}
g_2&=&-\frac{m_2}{\sqrt{m_1m_2}}\left[{\xi+\frac1{m_1}\rho_2+\frac1{m_2}\left(\rho_1+\xi_3\right)}\right],\\ 
\label{gexp}
  g&=&  \frac{1}{2\sqrt{m_1m_2}}\left[{\xi+\frac1{m_1}\rho_2 +\frac1{m_2}\left(\rho_1-\xi_3\right)}\right],\\
\label{a1exp}
a_1&=&-\frac{1}{\sqrt{m_1m_2}}\frac1{m_2}\frac1{\oh+1}\left[{-\bar\Lambda\xi+2(\oh+1)\chi_2-\xi_3}\right],\\
\label{a2exp}
a_2&=&-\frac{1}{\sqrt{m_1m_2}}\left[{\xi+\frac1{m_1}\left(\rho_1-\xi_3\right) +\frac1{m_2}\left(\rho_1-\xi_3\right)
-\frac{\bar\Lambda}{m_1}\frac1{\oh+1}\xi-\frac{2\chi_2}{m_1}+\frac1{m_1}\frac{\oh}{\oh+1}}\xi_3\right],\\
\label{fexp}
  f&=&{\sqrt{m_1m_2}}(\oh+1)\left[{\xi+\frac{\bar\Lambda\xi\oh}{\oh+1}\left(\frac1m_1+\frac1m_2\right)+\frac1m_1\rho_2
  +\frac1m_2\left(\rho_1-\frac{\oh-1}{\oh+1}\xi_3\right)    }\right]\\
\label{g0exp}
g_0&=&\frac{1}{(m_1m_2)^{3/2}}\left[\frac{\bla\xi+\xi_3}{\oh+1}+2\chi_2\right]  
\end{eqnarray}
For the following analysis it is worth noting that the behavior of the combination $2p_1p_2\cdot g-m_1\cdot g_2$ 
and $f$ in LO and NLO coincide, namely  
\begin{equation}
\label{fmodified}
f\simeq 2p_1p_2\cdot g-m_1\cdot g_2.
\end{equation}
It is also convenient to introduce the form factor $a'_2$ such that 
\begin{equation}
a'_2=a_2+2s.
\end{equation}
In what follows we need the expansions of the following linear combinations of the form factors $a'_2$ 
and $a_1$
\begin{eqnarray}
\label{aminus}
a_1m_2-a'_2m_1&=&-\frac1{\sqrt{m_1m_2}}\left[ 4\chi_2-\xi_3 \right],\\ 
\label{aplus}
a_1m_2+a'_2m_1&=&-\frac1{\sqrt{m_1m_2}}\left[ -\frac{2\bla\xi}{\oh+1}+\xi_3 \frac{\oh-1}{\oh+1} \right].
\end{eqnarray}
\section{Transition form factors in the dispersion quark model}
The results presented in the previous section are strict consequences of QCD in the heavy--quark limit 
which however cannot provide more information on the universal form factors $\xi$ and $\rho_1,\rho_2, \chi_2,\xi_3$. 
They must be calculated within a nonperturbative dynamical approach.
We study the form factors within the dispersion formulation of the quark model \cite{m1,m2} 
which has proved to be a reasonable framework for describing meson decays. 
We start with $q^2<0$ and represent the form
factors as double spectral representations in the invariant masses of the initial and final $q\bar q$ pairs. 
The form factors at $q^2>0$ are derived by performing the analytical continuation. 

\begin{figure}[hbt]
\begin{center}  
\mbox{\epsfig{file=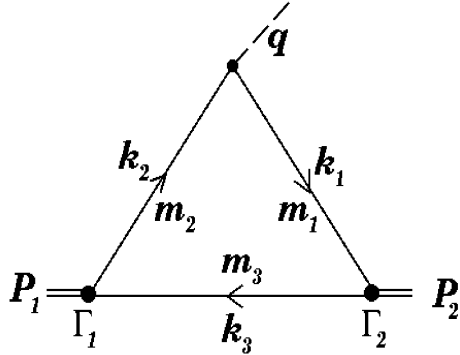,height=5.cm}  }
\end{center}
\caption{One-loop graph for a meson decay.\label{fig:trianglegraph}}
\end{figure}

The transition of the initial meson  $q(m_2)\bar q(m_3)$ with the mass $M_1$ 
to the final meson $q(m_1)\bar q(m_3)$ with the mass $M_2$ 
induced by the quark transition $m_2\to m_1$ through the current $\bar q(m_1) J_\mu q(m_2)$ is described by the 
diagram of Fig.\ref{fig:trianglegraph}. For constructing the double spectral representation we must  
consider a double--cut graph where all intermediate particles go on mass shell
but the initial and final mesons have the off--shell momenta 
$\tilde p_1$ and and $\tilde p_2$ such that $\tilde p_1^2=s_1$ and $\tilde p_2^2=s_2$ with 
$(\tilde p_1-\tilde p_2)^2\equiv s_3=q^2$ kept fixed. 

For the transition $B\to D,D^*$ $m_2=m_b, m_1=m_c$, and $m_3=m_u$. 
The constituent quark structure of the initial and final mesons is given in terms of the vertices $\Gamma_1$
and $\Gamma_2$, respectively. The initial pseudoscalar meson vertex has the spinorial structure
$\Gamma_1=i\gamma_5G_1/\sqrt{N_c}$; the final meson vertex has the structure 
$\Gamma_2=i\gamma_5G_2/\sqrt{N_c}$ for a pseudoscalar state and the structure 
$\Gamma_{2\mu}=[A\gamma_\mu+B(k_1-k_3)_\mu]\,G_2/\sqrt{N_c}$, 
$A=-1$, $B=1/(\sqrt{s_2}+m_1+m_3)$ for an $S$--wave vector meson. 

The double spectral densities $\tilde f$ of the form factors are obtained by calculating the relevant traces and
isolating the Lorentz structures depending on $\tilde p_1$ and $\tilde p_2$. 
The invariant factors of such Lorentz structures provide the double spectral densities $\tilde f$ corresponding 
to taking into account contributions of the two--particle singularities in the Feynman graph. 
Let us point out that this procedure allows one to obtain unsubtracted
double spectral densities, whereas subtraction terms should be determined independently. In this paper we
determine the subtraction terms from matching the $1/m_Q$ expanded form factors of the quark model to the
heavy--quark expansion in QCD. 

At $q^2<0$ the spectral representations of the form factors have the form \cite{m1}
\begin{equation}
\label{ff}
f_i(q^2)=
\frac1{16\pi^2}\int\limits^\infty_{(m_1+m_3)^2}ds_2\varphi_2(s_2)
\int\limits^{s_1^{+}(s_2,q^2)}_{s_1^{-}(s_2,q^2)}ds_1\varphi_1(s_1)
\frac{\tilde f_i(s_1,s_2,q^2)}{\lambda^{1/2}(s_1,s_2,q^2)},
\end{equation}
where the wave function $\varphi_i(s_i)=G_i(s_i)/(s_i-M_i^2)$ and 
$$
s_1^\pm(s_2,q^2)=
\frac{s_2(m_1^2+m_2^2-q^2)+q^2(m_1^2+m_3^2)-(m_1^2-m_2^2)(m_1^2-m_3^2)}{2m_1^2}
\pm\frac{\lambda^{1/2}(s_2,m_3^2,m_1^2)\lambda^{1/2}(q^2,m_1^2,m_2^2)}{2m_1^2}
$$
and 
$
\lambda(s_1,s_2,s_3)=(s_1+s_2-s_3)^2-4s_1s_2
$
is the triangle function. 

The unsubtracted double spectral densities $\tilde f_i(s_1,s_2,q^2)$ of the form factors read \cite{m1,m2}: 
\begin{eqnarray}
\label{s}
\tilde s&=&2\,[m_1\alpha_{2}+m_2\alpha_{1}+m_3(1-\alpha_{1}-\alpha_{2})],
\\
\label{f1}
\tilde f_1&=&2m_1\tilde s+4\alpha_2[s_2-(m_1-m_3)^2]-2m_3\tilde s, 
\\
\label{f2}
\tilde f_2&=&2m_2\tilde s+4\alpha_1[s_1-(m_2-m_3)^2]-2m_3\tilde s,
\\
\label{g}
\tilde g&=&A\tilde s-4B\beta,
\\
\label{g1}
\tilde g_{1}&=&A\tilde f_1-8\beta+8B(m_1+m_3)\beta,
\\
\label{g2}
\tilde g_{2}&=&A\tilde f_2+8B(m_2-m_3)\beta,
\\
\label{a2}
\tilde a_{2D}&=&-2\tilde s+ 4BC_2\alpha_1+\alpha_{12}C_0,
\\
\label{a1}
\tilde a_{1D}&=&-4A\,(2m_3+BC_1)\alpha_1+\alpha_{11}C_0,
\\
\label{f}
\tilde f_D&=&-4A[m_1m_2m_3+\frac{m_2}2(s_2-m_1^2-m_3^2)
+\frac{m_1}2(s_1-m_2^2-m_3^2)-\frac{m_3}2(s_3-m_1^2-m_2^2)]+C_0\beta,
\\
\label{g0}
\tilde g_{0D}&=&-8A\alpha_{12}-8B\,[-m_3\alpha_{1}+(m_3-m_2)\alpha_{11}+(m_3+m_1)\alpha_{12}],
\end{eqnarray}
where 
\begin{eqnarray}
\label{alpha1}
\alpha_1&=&\left[(s_1+s_2-s_3)(s_2-m_1^2+m_3^2)-2s_2(s_1-m_2^2+m_3^2)\right]/{\lambda(s_1,s_2,s_3)},
\\
\label{alpha2}
\alpha_2&=&
\left[(s_1+s_2-s_3)(s_1-m_2^2+m_3^2)-2s_1(s_2-m_1^2+m_3^2)\right]/{\lambda(s_1,s_2,s_3)},
\\
\label{beta}
\beta&=&\frac14\left[2m_3^2-\alpha_1(s_1-m_2^2+m_3^2)-\alpha_2(s_2-m_1^2+m_3^2)\right],
\\
\label{alpha11}
\alpha_{11}&=&\alpha_1^2+4\beta {s_2}/{\lambda(s_1,s_2,s_3)}, \quad 
\alpha_{12}=\alpha_1\alpha_2-2\beta(s_1+s_2-s_3)/\lambda(s_1,s_2,s_3),
\\
\label{cc}
C_0&=&-8A(m_2-m_3)+4BC_3,\quad C_1=s_2-(m_1+m_3)^2, 
\\
C_2&=&s_1-(m_2-m_3)^2, \quad C_3=s_3-(m_1+m_2)^2-C_1-C_2. \nonumber
\end{eqnarray}

We label with a subscript 'D' the double spectral densities of the form factors which 
require subtractions. 
We fix this subtraction procedure by requiring the $1/m_Q$ expansion of the form factors to have a 
proper form in accordance with QCD in leading and next--to--leading orders for the case 
of a meson transition caused by heavy--to--heavy quark
transition. As we shall see this yields the double spectral densities 
which include properly defined subtraction terms 
\begin{eqnarray}
\label{f+sub}
\tilde f&=&\tilde f_D+[(M_1^2-s_1)+(M_2^2-s_2)]\tilde g, \\
\label{a1+sub}
\tilde a_1&=&\tilde a_{1D}
+\frac{1}{(\oq+1)m_2}\left(\frac{M_1^2-s_1}{\sqrt{s_1}}+\frac{M_2^2-s_2}{\sqrt{s_2}}\right)
\frac{\tilde g}2,\\
\label{a2+sub}
\tilde a_2&=&\tilde a_{2D}
+\frac{1}{(\oq+1)m_1}\left(\frac{M_1^2-s_1}{\sqrt{s_1}}+\frac{M_2^2-s_2}{\sqrt{s_2}}\right)
\frac{\tilde g}2,\\
\label{g0+sub}
\tilde g_0&=&\tilde g_{0D}
+\frac{1}{(\oq+1)m_1m_2}\left(\frac{M_1^2-s_1}{\sqrt{s_1}}+\frac{M_2^2-s_2}{\sqrt{s_2}}\right)
\frac{\tilde g}2.
\end{eqnarray}

As the analytical continuation to the timelike region is performed, 
in addition to the normal contribution which is just the expression (\ref{ff}) taken
at $q^2>0$, anomalous contribution emerges \cite{bbd}. 
The origin of the anomalous contribution is connected with the motion of a zero of the 
triangle function $\lambda(s_1,s_2,s_3)$ from the unphysical sheet at $q^2<0$ onto the physical 
sheet at $q^2>0$ through the normal cut between the points $s_1^-(s_2,s_3)$ and $s_1^+(s_2,s_3)$.

To be more specific, we can write 
\beq
\lambda(s_1,s_2,s_3)=(s_1-s_1^L)(s_1-s_1^R),
\eeq
where $s_1^L=(\sqrt{s_2}-\sqrt{s_3})^2$ and 
$s_1^R=(\sqrt{s_2}+\sqrt{s_3})^2$. 
At $q^2<0$ both singularities $s_1^L$ and $s_1^R$ are located on the unphysical sheet and do not 
contribute to the spectral representation of the form factor. 
As $q^2$ becomes positive, $s_1^R$ moves around $s_1^-$ and appears on the physical sheet. 
The pinching of the points $s_1^R$ and $s_1^-$ occurs at $s_2^0$ such that 
$s_1^R(s_2^0)=s_1^-(s_2^0)$. The corresponding value $s_2^0(s_3)$ reads 
\begin{equation}
\sqrt{s_2^0}=-\frac{s_3+m_1^2-m_2^2}{2\sqrt{s_3}}+
\sqrt{
\left({  
\frac{s_3+m_1^2-m_2^2}{2\sqrt{s_3}}
}\right)^2+(m_3^2-m_1^2) 
},\quad s_3<(m_2-m_1)^2. 
\end{equation}
The spectral representation of the form factor at $q^2>0$ takes the form \cite{m1}
\beq
\label{final}
f(q^2)=\frac{1}{16\pi^2}
\int\limits_{(m_1+m_3)^2}^\infty\varphi_2(s_2)
\int\limits_{s_1^-}^{s_1^+}
\varphi_1(s_1)
\frac{\tilde f(s_1,s_2,q^2)}{\lambda^{1/2}(s_1,s_2,q^2)}
+\theta(q^2)\frac{1}{8\pi^2}\int\limits_{s_2^0}^\infty\varphi_2(s_2)
\int\limits_{s_1^R}^{s_1^-}
ds_1\varphi_1(s_1)\frac{\tilde f(s_1,s_2,q^2)}{\lambda^{1/2}(s_1,s_2,q^2)}. 
\eeq 
One should also take into account that if the double spectral density $f_i$ is singular at the point 
$s_1^R$ then a properly defined spectral representation contains also another kind of subtraction terms: 
these terms appear as the contribution of a small circle around $s_1^R$ which is the lower integration 
limit in the anomalous part. 
These subtractions have quite different nature and different form than subtractions  
in the spectral representations at $q^2<0$. The corresponding expression can be found in \cite{m2}.

The normal contribution dominates the form factor at small timelike and vanishes as 
$q^2=(m_2-m_1)^2$ while the anomalous contribution is negligible at small $q^2$ and steeply rises
as $q^2\to(m_2-m_1)^2$. 
 
For pseudoscalar and vector mesons with the mass $M$ built up of the 
constituent quarks $m_q$ and $m_{\bar q}$, the function $\varphi$ is normalized as follows \cite{m1}
\begin{equation}
\label{norma}
\frac1{8\pi^2}\int ds \varphi^2(s)\frac{\lambda^{1/2}(s,m_q^2,m_{\bar q}^2)}{s}
[s-(m_q-m_{\bar q})^2]=1. 
\end{equation}
This equation is the normalization of the elastic charge  form factor at $q^2=0$. 

The meson wave function can be written in the form 
\begin{equation}
\label{5vertex}
\varphi(s)=\frac{\pi}{\sqrt{2}}\frac{\sqrt{s^2-(m_1^2-m_2^2)^2}}
{\sqrt{s-(m_1-m_2)^2}}\frac{1}{s^{3/4}}w(k),\qquad
k=\frac{\lambda^{1/2}(s,m_1^2,m_2^2)}{2\sqrt{s}}
\end{equation}
where $w(k)$ is the ground--state $S$--wave radial wave function. 

In the next sections we analyze the form factors given by the dispersion representation 
(\ref{final}) with the spectral densities (\ref{s}--\ref{g2}) and (\ref{f+sub}--\ref{g0+sub}) 
and demonstrate them to have the following properties in the case of a heavy parent meson: 
for the transition induced 
by the heavy--to--heavy quark transition they satisfy the LO and NLO relations \cite{luke} 
of the $1/m_Q$ expansion in accordance with QCD 
provided the functions $\varphi_i$ are localized near the $q\bar q$ threshold with the width of order
$\Lambda_{QCD}$. For the meson decay induced by the heavy--to--light quark transition they satisfy 
the LO relations between the form factors of the vector, axial--vector and tensor currents \cite{iwhl}.  
\section{Heavy--quark expansion in quark model for heavy--to--heavy transitions}
In this section we consider the form factors of the dispersion quark model in 
the case when both $m_2$ and $m_1$ are large. We calculate the universal form factors and 
demonstrate that requiring the structure of the $1/m_Q$ expansion in 
the quark model to be consistent with the structure of such expansion in QCD allows us to determine the subtraction 
terms. 
 
\subsection{Soft wave function and normalization condition} 
First, we need to specify the properties of the soft wave function of a heavy meson. 
A basic property of such soft wave function $\varphi(s,m_Q,m_{\bar q},\bla)$ is a strong peaking near the $q\bar q$ 
threshold. 
For elaborating the $1/m_Q$ expansion, it is convenient to formulate such peaking in terms of the 
variable $z$ such that $s=(m_Q+m_3+z)^2$ (hereafter we denote the mass of the light quark as $m_3$).
The region above the $q\bar q$ threshold which contributes to the 
spectral representation corresponds to $z>0$. A localization of the soft wave function in terms of $z$ means that the 
wave function is nonzero as $z\le \Lambda_{QCD}$. In the heavy meson case $m_Q \gg m_{3}\simeq z\simeq \bla$.
Let us notice that for a heavy meson the localization in terms of $z$ is equivalent to the localization in terms of the 
relative momentum in the meson rest frame 
\begin{equation}
\label{z2k}
{\vec k}^2=z(z+2m_3)+O(1/m_Q).
\end{equation}
The normalization condition (\ref{norma}) which is 
a consequence of the vector current conservation in the full theory
provides an (infinite) chain of relations in the effective theory.  
Namely, expanding the soft wave function in $1/m_Q$ as follows 
\begin{equation}
\label{phi}
\varphi(s,m_Q,m_{3},\bla)=\frac\pi{\sqrt{m_Q}}\phi_0(z,m_{3},\bla)\left[{1+\frac {m_3}{4m_Q}
\chi_1(z,m_{3},\bla)+O(1/m_Q^2)}\right],
\end{equation}
we come to the normalization condition in the form 
\begin{equation}
\int dz \phi^2_0(z)\sqrt{z}(z+2m_3)^{3/2}\left[1+\frac{m_3}{2m_Q}\chi_1(z)-\frac{m_3}{2m_Q}+\ldots\right]=1. 
\end{equation}
This exact relation is equivalent to an infinite chain of equations in different $1/m_Q$ orders. 
Lowest order relations take the form  
\begin{eqnarray}
\label{normaphi}
&\int &dz\;\phi^2_0(z)\sqrt{z}(z+2m_3)^{3/2}=1, \\
&\int & dz\;\phi^2_0(z)\sqrt{z}(z+2m_3)^{3/2}\chi_1(z)=1. 
\end{eqnarray}

\subsection{The variables $\oh$ and $\oq$}
In the description of the transition processes the dispersion formulation 
of the quark model has the following feature: 
since the underlying process is the quark transition, the relevant kinematical variable for the description of the
dynamics of the process is the quark recoil $\oq$ which is defined as follows
\beq
q^2=(m_2-m_1)^2-2m_1m_2(\oq-1).
\eeq
The relationship between $\oh$ and $\oq$ is given by the condition that the spectral representation for the
form factor is written at fixed value of $q^2$ and hence 
\beq
q^2=(M_1-M_2)^2-2M_1M_2(\oh-1)=(m_2-m_1)^2-2m_1m_2(\oq-1).
\eeq
In the case of heavy particle transitions these quantities are related to each other as
\beq
\oq=\oh+\bla\left(\frac{1}{m_1}+\frac{1}{m_2}\right)(\oh-1)+O(1/m_Q^2).
\eeq
We shall obtain the representations of the form factors as functions of the variable $\oq$. 
The variables $\oh$ and $\oq$ are different by the terms of order $1/m_Q$ at nonzero recoil. 
On the other hand, the quark and meson zero recoil points coincide with $1/m_Q^2$ accuracy. 
This means that in the analyses of the $1/m_Q$ expansion at nonzero recoil 
the difference between the $\oh$ and $\oq$ might be ignored in the Isgur--Wise function, but gives nontrivial
contribution to the NLO form factors. At the same time, at zero recoil the difference between $\oh$ and $\oq$ 
might be neglected both in the leading and next--to--leading orders. 
Namely, the quark model provides the expansion of the form factor in the following form 
\beeq
h&=&h_0(\oq)+\frac{1}{m_1}h^{(1)}_1(\oq)+\frac{1}{m_2} h^{(2)}_1(\oq)+\ldots=\\
 &=&h_0(\oh)+h'_0(\oh)\bla\left(\frac{1}{m_1}+\frac{1}{m_2}\right)(\oh-1)
   +\frac{1}{m_1}h^{(1)}_1(\oq)+\frac{1}{m_2}\bar h^{(2)}_1(\oq)+\ldots.
\eeeq
As we shall see later, among the NLO form factors only $\rho_{1,2}$ are affected by the 
the difference between $\oh$ and $\oq$ whereas $\xi, \xi_3, \chi_2$ are not. 

\subsection{Relative magnitudes of the normal and the anomalous contributions}
We are going now to demonstrate that the anomalous contribution comes into the game only in close vicinity 
of the zero recoil point whereas beyond this region is negligible. 

Let us study the behavior of the anomalous contribution in the region  
\beq 
\oq-1\simeq m_Q^{-(2+\eps)}.
\eeq
Introducing the variables $z_1$ and $z_2$ such that $s_1=(m_2+m_3+z_1)^2$ and $s_2=(m_1+m_3+z_2)^2$ we find that 
the magnitude of the anomalous contribution is controlled by the value of $z_2^0(\oq)$ such that 
$s_2^0=(m_1+m_3+z_2^0(\oq))^2$ 
which is the lower boundary of the $z_2$ integration. If $z_2^0(\oq)$ becomes large, i.e. of the order $m_Q$, 
the anomalouis contribution is suppressed by the fall--down of the soft wave function.
This suppression is at least stronger than $1/m_Q^2$. This means that the anomalous contribution is nonvanishing only if  
\beq
z_2^0(\oq)=\frac{m_1m_2\sqrt{\oq^2-1}+m_1m_2\oq-m_1^2}{\sqrt{m_1^2+m_2^2-2m_1m_2\oq}}-m_1+O(m_3)\simeq \bla.
\eeq
In the region $\oq-1\simeq m_Q^{-(2+\eps)}$, one finds $z_2^0(\oq)\simeq m_Q^{-\eps/2}$. 
Hence the anomalous contribution comes actually into the game only in the $O(1/m_Q^2)$ vicinity of the 
zero recoil point but otherwise might be neglected. On the other hand, at the quark zero recoil point $\oq=1$, 
the normal contribution vanishes and the form factor is given by the anomalous contribution. 

We shall calculate the form factors in the region $\oq-1=O(1)$ where only the normal contribution should be 
taken into account in leading and subleading orders. 

\subsection{The LO analysis}
To perform the LO analysis of the form factors let us start with the integration measure. 
With $1/m_Q$ accuracy it can be represented in the form 
\beq
\frac1{16\pi^2}\frac{ds_1ds_2\theta\left(s_2\ge(m_1+m_3)^2\right)\theta(s_1^-\le s_1\le s_1^+)}
{\lambda^{1/2}(s_1,s_2,s_3)}
\simeq
\frac1{4\pi^2}dz_2\sqrt{z_2(z_2+2m_3)}\frac{d\eta}{2}\theta(z_2)\theta(\eta^2<1), 
\eeq
and the expression for $z_1$ reads 
\beq
\label{z1}
z_1=z_2\oq +m_3(\oq -1)+\eta\sqrt{z_2(z_2+2m_3)}\sqrt{\oq ^2-1}+O(1/m_Q).
\eeq
Let us point out that the LO integration measure is symmetric in $z_1$ and $z_2$. 

Next, we need expanding the spectral densities (\ref{s}--\ref{g0}). To this end we must take 
into account that under the integral sign 
$z_1$ and $z_2$ are localized in the region $z\le\bla$ due to the 
soft wave functions $\phi(z)$. 

In LO the kinematical coefficients (\ref{alpha1}--\ref{cc}) in the region $\oq-1=O(1)$ simplify to 
\beeq
\lambda(s_1,s_2,q^2)&=&4m_1^2m_2^2(\oq^2-1),\\
\alpha_1&=&\frac1{m_2(\oq+1)}\left[ m_3+z_2+\frac{z_2-z_1}{\oq-1}\right],\\
\alpha_2&=&\frac1{m_1(\oq+1)}\left[ m_3+z_1+\frac{z_1-z_2}{\oq-1}\right],\\
\beta&=&\frac12\left[m_3^2-\frac2{\oh+1}(m_3+z_1)(m_3+z_2)+ \frac{(z_1-z_2)^2}{\oq^2-1}\right],\\
\alpha_{11}&=&\alpha_1^2+\frac\beta{m_2^2(\oq^2-1)},\quad 
\alpha_{12}=\alpha_1\alpha_2-\frac{\beta\;\oq}{m_1m_2(\oq^2-1)},\\
C_0&=&-4m_2(\oq-1),\quad B=\frac1{2m_1}, \quad C_1=2m_1z_2,\quad C_2=2m_2(z_1+2m_3).  
\eeeq

One finds the LO behavior of the form factor densities (\ref{s}--\ref{a2}) is determined by the term proportional to
$\tilde s$. The latter reads in the LO 
\beq
\tilde s\simeq 2\left( m_3+\frac{z_1+z_2+2m_3}{\oq+1} \right). 
\eeq
The LO expression for $\tilde f$ takes the form  
\beq
\tilde f_D=(\oq+1)\left(m_3+\frac{z_1+z_2+2m_3}{\oq+1}\right).
\eeq
The spectral densities $\tilde a_1$ and $\tilde g_{0D}$ vanish in the leading order. 

Hence the LO relations (\ref{f1exp}--\ref{g0exp}) are fulfilled with the Isgur--Wise (IW) function 
\begin{eqnarray}
\label{xi}
\xi(\oh )=\int
dz_2 \phi_0(z_2)\sqrt{z_2(z_2+2m_3)}\int\limits_{-1}^{1}\frac{d\eta}2
\phi_0(z_1)\left({m_3+\frac{2m_3+z_1+z_2}{1+\oh }}\right). 
\end{eqnarray}
In (\ref{xi}) we used the equality of $\oq$ and $\oh$ with $1/m_Q$ accuracy. 
The normalization condition of the LO wave functions (\ref{normaphi}) yields $\xi(1)=1$. 
For the slope of the IW function at zero recoil, $\rho^2=-\xi'(1)$, one finds 
\begin{eqnarray}
\label{rho2}
\rho^2=\frac13\int
dz \sqrt{z}(z+2m_3)^{3/2}{\left(\phi_0'(z)\right)}^2z(z+2m_3)
\end{eqnarray}
 
Let us point out that the subtraction terms in the spectral densities do not contribute in the LO 
relations. As we shall see later, they are important in the NLO analysis. 

\subsection{The NLO analysis of the form factors $s$, $f_1$, $f_2$, $g_1$, $g_2$, and $g$ }

First, let us concentrate on the NLO relations (\ref{f1exp}--\ref{gexp}). 
It is convenient to analyze the linear combinations of the form factors 
which do not contain the LO contribution. These combinations are 
\begin{eqnarray}
\label{1exp}
g-s&=&\frac{1}{2\sqrt{m_1m_2}}\frac{1}{m_1}\left[\rho_2-\rho_1+\xi_3\right],\\ 
\label{2exp}
g_1+f_1&=&\frac{1}{\sqrt{m_1m_2}}\left[-\rho_2+\rho_1+\xi_3\right],\\
\label{3exp}
g_2+f_2&=&-\frac{m_2}{m_1}\frac{1}{\sqrt{m_1m_2}}\left[\rho_2-\rho_1+\xi_3\right],\\
\label{4exp}
f_1-2m_1s&=&\frac{2\xi_3}{\sqrt{m_1m_2}}. 
\end{eqnarray}

The spectral densities of the form factor combinations in the l.h.s. of eqs. (\ref{1exp}--\ref{3exp}) read  
\begin{eqnarray}
\label{1dexp}
\tilde g-\tilde s&=&-\frac{2}{m_1}\beta,\\ 
\label{2dexp}
\tilde g_1+\tilde f_1&=&-4\beta,\\
\label{3dexp}
\tilde g_2+\tilde f_2&=&4\frac{m_2}{m_1}\beta,
\end{eqnarray}
Comparison with the eqs. (\ref{1exp}--\ref{3exp}) yield the relation 
\beq
\rho_1(\oh)=\rho_2(\oh), 
\eeq
For the form factor $\xi_3$ we come to the representation 
\begin{eqnarray}
\label{xi3a}
\xi_3(\oh )=-\int 
dz_2 \phi_0(z_2)\sqrt{z_2(z_2+2m_3)}\int\limits_{-1}^{1}\frac{d\eta}2\phi_0(z_1) 
\frac12\left[m_3^2-\frac2{\oh+1}(m_3+z_1)(m_3+z_2)+ \frac{(z_1-z_2)^2}{\oh^2-1}\right], 
\end{eqnarray}
with $z_1$ given by the expression (\ref{z1}). In (\ref{xi3a}) we have neglected the $O(1/m_Q)$ difference between 
$\oh$ and $\oq$. 

On the other hand, the equation (\ref{4exp}) yields the representation for the form factor $\xi_3$ in a different
form 
\begin{eqnarray}
\label{xi3b}
\xi_3(\oh)&=&\int 
dz_2 \phi_0(z_2)\sqrt{z_2(z_2+2m_3)}\int\limits_{-1}^{1}\frac{d\eta}2\phi_0(z_1)\nonumber \\
&\times&\left[\frac{z_2+2m_3}{\oh+1}\left(z_2+m_3+\frac{(z_1-z_2)\;\oh}{\oh-1}\right)
-\frac{m_3}{2}\left({m_3+\frac{2m_3+z_1+z_2}{1+\oh }}\right)\right]. 
\end{eqnarray}
One can check that for the soft wave functions providing convergency of the integrals and nonsingular at $z=0$ 
the representations (\ref{xi3a}) and (\ref{xi3b}) are equivalent. 
At zero recoil one finds 
\beq
\xi_3(1)=\int dz_2\;\sqrt{z_2}(z_2+2m_3)^{3/2}\;\phi^2_0(z_2)\frac{z_2}3\equiv\frac{\langle{z}\rangle}{3}, 
\eeq
where $\langle z \rangle$ is an average kinetic energy of the light constituent quark inside an (infinitely) 
heavy meson in its rest frame.  
It is worth noting that the function $\xi_3$ is positive for all $\oh$. 

The universal form factor $\rho_1=\rho_2$ can be found from the expansion of $s(\oh)$ with a NLO accuracy. 
In this case the
NLO terms in the $1/m_Q$ expansions of the integration measure, the wave function, and the spectral density $\tilde s$
contribute. 
Namely, we can write for the form factor $h_s$ the expression 
\beeq
h_s(\oh)&=&\int\left[d\mu_0+\frac{d\mu_1^{(1)}}{m_1}+\frac{d\mu_1^{(2)}}{m_2}\right]
\left[\tilde s(\oq)+\frac{S_1}{m_1}
+\frac{S_2}{m_2}\right]\phi_0(z_1)
\left(1+\frac{m_3}{4m_1}\chi_1(z_1)\right) \phi_0(z_2)
\left(1+\frac{m_3}{4m_2}\chi_1(z_2)\right)\nonumber\\
&=&\xi(\oq)+\left( \frac1{m_1} + \frac1{m_2}\right)\xi^{(1)}(\oq)+\ldots
=\xi(\oh)+\left( \frac1{m_1} + \frac1{m_2}\right)[\xi^{(1)}(\oh)+\xi'(\oh)\bla(\oh-1)]+\ldots,
\eeeq
and hence $\rho_1(\oh)=\xi^{(1)}(\oh)+\xi'(\oh)\bla(\oh-1)-\xi_3(\oh)-\bl2\xi(\oh)$. 
We do not present explicit expression for $\rho_1$. However, let us consider $\rho_1$ at zero recoil. 
The analysis of the anomalous contribution at $\oq=1$ gives
\beq
h_s(\oq=1)=1+\left( \frac1{m_1} + \frac1{m_2}\right) \left( \bl2-\frac{\langle z \rangle}{3}\right)+\ldots
\eeq
Using the relations $h_s(\oh=1)=h_s(\oq=1)+O(1/m_Q^2)$ and $\xi_3(1)=\langle z \rangle/3$ 
we obtain $\rho_1(1)=0$ just as required by the HQET.  

\subsection{The NLO analysis of the form factor $f$ }
First, let us demonstrate that the dispersion representation of the form factor $f$ requires subtraction.
To this end consider the anomalous contribution at $\oq=1$. For the form factor $f_D$ 
constructed from the spectral density $\tilde f_D$ through the unsubtracted double dispersion representation we find
\footnote{Hereafter we denote
as $f_{\tilde a}$ the form factor constructed from the spectral density $\tilde a$ 
through unsubtracted double dispersion representation.}
\beq
f_D(1)=2\sqrt{M_1M_2}\left[1+\frac12\left(\frac1{m_1}+\frac1{m_2} \right)
\left(\langle z \rangle+m_3-\bla\right)+\ldots \right]
\eeq
This value contradicts the Luke theorem which requires the $1/m_Q$ corrections to vanish at zero recoil.
Let us demonstrate that the form factor $f$ constructed from the 
spectral density with the included subtraction term
\beq
\label{tildeffull}
\tilde f=\tilde f_D+(M_1^2-s_1+M_2^2-s_2)\tilde g=\tilde f_D+(2p_1p_2-2\tilde p_1 \tilde p_2)\tilde g, 
\eeq
satisfies the NLO relation (\ref{fexp}).
Here $2\tilde p_1 \tilde p_2= s_1+s_2-s_3$. 

We have noticed above that to LO and NLO the expansion of the form factor $f$ and the combination 
$2p_1p_2\cdot g-m_1\cdot g_2$ coincide (\ref{fmodified}). 
Hence checking the NLO relations for the form factor $f$ is equivalent to checking 
with the NLO accuracy the relation 
\beq
\label{fnew}
f_D\simeq f_{2\tilde p_1\tilde p_2\cdot \tilde g} -m_1 \cdot g_2. 
\eeq
The spectral density of the r.h.s. of eq. (\ref{fnew}) can be written as
\beq
\label{fsym}
2\tilde p_1\tilde p_2\cdot \tilde g -m_1 \cdot \tilde g_2\simeq 
2m_1m_2(\oq+1)\tilde s+ [2m_1(z_2+m_3)+2m_2(z_1+m_3)]\tilde s-8(m_1+m_2)\beta-4m_2(\oq-1)\beta. 
\eeq
For checking the expression (\ref{fnew}) in NLO we need the expansion of the spectral density $\tilde s$ in LO and NLO
which has the structure
\beq
\frac12\tilde s=m_1\alpha_2+ m_2\alpha_1+m_3(1-\alpha_1-\alpha_2)\equiv m_3+\frac{z_1+z_2+2m_3}{\oq+1}
+\frac{S_1}{m_1}+\frac{S_2}{m_2}
+\ldots,
\eeq
with 
\beq
S_1=\frac12 z_2(z_2+2m_3)-z_1m_3-(z_2+2m_3)\frac{z_1-z_2}{\oq-1}-\frac{(z_1+z_2+2m_3)^2}{\oq+1}, 
\eeq
and $S_2$ is obtained from $S_1$ by replacing  $z_1$ and $z_2$. 
The spectral density $\tilde f_D$ reads
\beq
\tilde f_D=4m_1m_2(\oq+1)\left(m_3+\frac{z_1+z_2+2m_3}{\oq+1}\right)
+2m_2z_2(z_2+2m_3)+2m_1z_1(z_1+2m_3)-4m_2(\oq-1)\beta.
\eeq
Notice that for checking the NLO relation (\ref{fnew}) between the form factors 
we do not need explicit expression for the integration measure in the NLO: 
in the LO the spectral densities are equal 
and hence the NLO contributions from the integration measure into both sides of the eq (\ref{fnew}) are equal too. 
Finally, the eq. (\ref{fnew}) is satisfied if the following relation is valid
\begin{eqnarray}
\label{fconsisitency}
&\int& dz_2\;\sqrt{z_2(z_2+2m_3)}\phi_0(z_2)\int\limits_{-1}^{1}\frac{d\eta}2\phi_0(z_1)\nonumber \\
&&\times\left[
-z_2(z_2+2m_3)+2(z_1+m_3)\left( m_3+\frac{z_1+z_2+2m_3}{\oq+1}  \right)-4\beta +2(\oq+1)S_1\right]=0,
\end{eqnarray}
with $z_1$ given by (\ref{z1}). One can check this relation to be true for any 
function $\phi_0(z)$ regular at $z=0$. Hence, the form factor $f$ calculated with the 
subtracted double spectral density 
(\ref{tildeffull}) satisfies the HQET relations in LO and NLO at all $\oh$.  

Strictly speaking, the NLO analysis does not allow us to uniquely specify the subtraction term: namely,
any spectral density of the form 
\beq
\tilde f=\tilde f_D +(M_1^2-s_1+M_2^2-s_2) \tilde \rho_f 
\eeq
has a proper NLO behavior in accordance with (\ref{fexp}) provided the spectral density $\tilde \rho_f$ 
behaves in the LO as 
$$
\tilde \rho_f\simeq 2\left(m_3+\frac{z_1+z_2+2m_3}{\oq+1}\right).
$$
As we demonstrate in the next section, the analysis of the heavy--to--light LO relations requires the 
identification 
$\tilde\rho_f=\tilde g.$  

Let us notice that the form factor obtained within the light--cone approach \cite{jaus} can be represented 
as dispersion representation with the spectral density \cite{m2}
\beq
\tilde f_{LC}=\frac{M_2}{\sqrt{s_2}}\tilde f_{D}+
M_2\left({\frac{s_1-s_2-s_3}{2\sqrt{s_2}}-\frac{M_1^2-M_2^2-s_3}{2M_2}}\right)
\frac{\tilde a_1+\tilde a_2}2. 
\eeq
At zero recoil one finds 
\beq
f_{LC}(1)=2\sqrt{M_1M_2}\left[1-\frac12\left(\frac1{m_1}-\frac1{m_2} \right)
\left(\langle z \rangle+m_3-\bla
\right)+\ldots \right]
\eeq
that contradicts the Luke theorem. 

\subsection{The NLO analysis of the form factors $a_{1,2}$ and $\chi_2$}
The unsubtracted spectral densities of the form factors $a_1$ and $a_2'\equiv a_2-2s$ in the LO read 
\beeq
\tilde a_{1D}&=&4\left[(z_2+2m_3)\alpha_1-m_2\alpha_{11}(\oq-1)\right],\\
\tilde a'_{2D}&=&4\left[\frac{m_2}{m_1}(z_2+2m_3)\alpha_1-m_2\alpha_{12}(\oq-1)\right]. 
\eeeq
The quantity $a_2'$ is more convenient than $a_2$ for calculations as 
the LO term of the heavy quark expansion of $a_2'$ is zero. 

The unsubtracted spectral representation for 
$a_1(\oh)m_2-a_2(\oh)m_1$ in combination with eq (\ref{aminus}) gives 
\beq
\chi_2(\oh)=\xi_3(\oh)-\frac14
\int
dz_2 \phi_0(z_2)\sqrt{z_2(z_2+2m_3)}\int\limits_{-1}^{1}\frac{d\eta}2
\phi_0(z_1)\left[\tilde a_1m_2-\tilde a_2m_1\right]. 
\eeq
Substituting the representation (\ref{xi3a}) for $\xi_3$ we find 
\beq
\chi_2(\oh)=0.
\eeq

Let us now consider the linear combination $a_1(\oh)m_2+a_2'(\oh)m_1$. As a first step, show that the unsubtracted 
dispersion representation is not compatible with HQET. To this end calculate the unsubtracted form factors 
$a_{1D}$ and $a_{2D}'$ at zero recoil: 
\beeq
a_{1D}(1)&=&\frac1{\sqrt{m_1m_2}}\frac1m_2\left[ \frac{2}{3}\langle z \rangle+\frac {m_3}{2}  \right],\\
a'_{2D}(1)&=&\frac1{\sqrt{m_1m_2}}\frac1m_1\left[ \frac{1}{3}\langle z \rangle+\frac {m_3}{2}  \right],
\eeeq
and hence 
\beq
a_{1D}(1)m_2+a'_{2D}(1)m_1=\frac1{\sqrt{m_1m_2}}[\langle z \rangle+{m_3}]
\eeq
in contradiction with the HQET result eq. (\ref{aplus}) 
\beq
\frac{\bla}{\sqrt{m_1m_2}}.
\eeq
This fact suggests a necessity of subtraction in the quantity $a_1m_2+a_2'm_1$. 
Let us write the spectral density with subtraction in the form 
\beq
\tilde a_{1}m_2+\tilde a'_{2}m_1=\tilde a_{1D}m_2+\tilde a'_{2D}m_1+\frac\kappa{\oq+1}
\left(\frac{M_1^2-s_1}{2\sqrt{s_1}}+\frac{M_2^2-s_2}{2\sqrt{s_2}}\right)\frac{\tilde \rho_a}2
\eeq
with 
$$
\tilde \rho_a\simeq 2\left(m_3+\frac{z_1+z_2+2m_3}{\oq+1}\right).
$$

Then corresponding representation for the form factor reads 
\beeq
&a_1(\oh)m_2+a_2'(\oh)m_1=
\frac1{4\sqrt{m_1m_2}}
\int
dz_2 \phi_0(z_2)\sqrt{z_2(z_2+2m_3)}\int\limits_{-1}^{1}\frac{d\eta}2\phi_0(z_1)\\
&\times\left[ \tilde a_{1D}m_2+\tilde a'_{2D}m_1-\frac{2\kappa}{\oh+1}(z_1+z_2+2m_3)
\left(m_3+\frac{z_1+z_2+2m_3}{\oh+1}    \right)     
\right]+\frac{\kappa\bla}{2\sqrt{m_1m_2}}\frac{\xi(\oh)}{\oh+1}.
\eeeq
According to the HQET eq (\ref{aplus}) this relation should be equal to 
\beq
\frac{1}{\sqrt{m_1m_2}}\left[\frac{2\bla}{\oh+1}\xi(\oh) -\xi_3(\oh)\frac{\oh-1}{\oh+1}  \right]
\eeq
The term proportional $\xi(\oh)$ yields $\kappa=4$. One can check that this value also makes 
other parts of both expressions equal. 
Hence we arrive at the subtracted spectral density 
\beq
\tilde a_{1}m_2+\tilde a'_{2}m_1=\tilde a_{1D}m_2+\tilde a'_{2D}m_1+\frac{1}{\oq+1}
\left(\frac{M_1^2-s_1}{\sqrt{s_1}}+\frac{M_2^2-s_2}{\sqrt{s_2}}\right)\tilde \rho_a.
\eeq
The resulting spectral densities of the form factors $a_1$ and $a_2$ with the built--in subtraction terms take 
the form 
\beeq
\tilde a_1&=&\tilde a_{1D}
+\frac{1}{(\oq+1)m_2}\left(\frac{M_1^2-s_1}{\sqrt{s_1}}+\frac{M_2^2-s_2}{\sqrt{s_2}}\right)
\frac{\tilde \rho_a}2,\\
\tilde a_2&=&\tilde a_{2D}
+\frac{1}{(\oq+1)m_1}\left(\frac{M_1^2-s_1}{\sqrt{s_1}}+\frac{M_2^2-s_2}{\sqrt{s_2}}\right)
\frac{\tilde \rho_a}2. 
\eeeq

\subsection{The NLO analysis of $g_0$.}
The unsubtracted spectral density $\tilde g_{0D}$ has the form 
\beq
\tilde g_{0D}\simeq\frac4{\sqrt{m_1m_2}}\left[m_2\alpha_1(m_3+m_2\alpha_1+m_1\alpha_2)-\frac\beta{\oq+1}\right].
\eeq
At zero recoil one finds
\beq
g_{0D}(1)=\frac{1}{(m_1m_2)^{3/2}}\left[ \frac{\langle z \rangle+m_3}{2}+\frac{\langle z \rangle}6\right].
\eeq
On the other hand, taking into account our earlier finding $\chi_2=0$, the HQET result (\ref{g0exp}) reads 
\beq
g_{0D}(1)=\frac{1}{(m_1m_2)^{3/2}}\left[ \frac{\bla}{2}+\frac{\langle z \rangle}6\right].
\eeq
Our experience obtained in considering $a_{1}m_2+a'_2m_1$ hints that the subtraction procedure adds a term
proportional at zero recoil to $\bla-\langle z \rangle-m_3$, and hence we expect subtraction to work properly also
in the case of $g_0$. 

As a matter of fact, the spectral density 
\beq
\label{tildegsub}
\tilde g_0=\tilde g_{0D}+\frac{1}{(\oq+1)m_1m_2}
\left(\frac{M_1^2-s_1}{\sqrt{s_1}}+\frac{M_2^2-s_2}{\sqrt{s_2}}\right)\frac{\tilde \rho_{g_0}}2
\eeq
satisfies the HQ expansion (\ref{g0exp})
provided the function $\tilde \rho_{g_0}$ behaves in the LO as 
$$
\tilde \rho_{g_0}\simeq 2\left(m_3+\frac{z_1+z_2+2m_3}{\oq+1}\right).
$$

Concluding this section let us summarize our main results: we have calculated the universal form factors
and demonstrated the dispersion representations with relevant subtractions in the case of 
heavy--to--heavy transitions to reproduce the structure of the heavy--quark expansion in QCD in the leading 
and next--to--leading orders. 
However, we have not been able to fix uniquely these subtractions. As we shall see in the next section, 
the heavy--to--light transitions provide further
restrictions on the form of the subtraction terms. 
\section{Heavy--to--light meson transitions}
In this section we discuss meson decays induced by the heavy--to--light quark transitions in which case 
$M_1=m_2+O(1)$ is large, while $M_2\simeq m_1$ is kept finite. 
As found by Isgur and Wise \cite{iwhl}, in the region $\oh-1=O(1)$ 
the form factors of the tensor current can be expressed through the form factors of the vector and 
axial--vector currents in the leading $1/m_2$ order as follows 
\begin{eqnarray}
\label{hls}
s(q^2)&=&\frac{1}{2M_1}f_1(q^2),  \\
\label{hlg2}
g_{2}(q^2)&=&-2M_1\,g(q^2), \\
\label{hlg0}
g_0(q^2)&=&\frac{1}{M_1}[2g(q^2)+a_{2}(q^2)], \\
\label{hlg1}
g_{1}(q^2)&=&\frac{1}{M_1}[-f(q^2)+2p_1p_2\cdot g(q^2)]. 
\end{eqnarray}
 
We address the two issues: 
(i) perform the leading order $1/m_2$ expansion of the form factors and show the fulfillment of the relations
(\ref{hls}--\ref{hlg1}) and 
(ii) discuss the scaling behavior of the 
form factors of the transition of different heavy mesons into a fixed final light state. 

\subsection{The LO $1/m_Q$ expansion of the form factors in the quark model}
In the case of heavy--light transitions one observes the appearance of the two scales: 
the light scale $M_2\simeq m_1\simeq m_3 \simeq \bla$, and the heavy scale $M_2\simeq m_1$, and we may 
expand the form factors in inverse powers of the small parameter $\bla/m_2$. 
The kinematical coefficients in the leading $\bla/m_2$ order in the region $\oq-1=O(1)$ simplify to
\beeq
\label{hlcoefficients}
\lambda(s_1,s_2,q^2)&\simeq &4m_2^2\left[(z_1+m_3+\oq m_1)^2-s_2 \right],\quad \beta=O(1),\\
\alpha_1&\simeq&\frac{1}{m_2}
\frac{(z_1+m_3+\oq m_1)(s_2-m_1^2+m_3^2)/2-s_2(z_1+m_3)}
{(z_1+m_3+\oq m_1)^2-s_2}=O\left(\frac{1}{m_2}\right),\\
\alpha_2 &\simeq &
\frac{(z_1+m_3+\oq m_1)(z_1+m_3)-(s_2-m_1^2+m_3^2)/2}
{(z_1+m_3+\oq m_1)^2-s_2}=O\left(1\right),\\
\alpha_{12} &\simeq &\alpha_1\alpha_2-\frac{\beta(z_1+m_3+\oq m_1)}{2m_2\left[(z_1+m_3+\oq m_1)^2-s_2
\right]}=O\left(\frac{1}{m_2}\right),\\
\alpha_{11} &\simeq  &\alpha_1^2+\frac{1}{m_2^2}\frac{\beta s_2}{\left[(z_1+m_3+\oq
m_1)^2-s_2\right]}=O\left(\frac{1}{m_2^2}\right).
\eeeq
For the double spectral densities $\tilde f_i$ these expressions yield the following LO relations
\beeq
\label{hlds}
\tilde s&=&\frac{1}{2m_2}\tilde f_2, \\
\label{hldg2}
\tilde g_{2}&=&2m_2\tilde g,\\
\label{hldg0}
\tilde g_{0D}&=&
\frac{1}{m_2}(2\tilde g+\tilde a_{2D}), \\
\label{hldg1}
\tilde g_{1}&=&\frac{1}{m_2}[-\tilde f_D+2\tilde p_1\tilde p_2\cdot\tilde g],
\eeeq
with $2\tilde p_1\tilde p_2=s_1+s_2-s_3$. 
Notice that these relations hold also for $\oq=O(m_2)$. 

First two equations directly give the LO equality of the corresponding form factors. 
The relation (\ref{hldg0}) between the unsubtracted spectral densities is more interesting and requires 
the LO identity of the subtraction terms 
\beq
\tilde \rho_a\simeq \tilde \rho_{g_0}.
\eeq
The choice $\tilde \rho_a=\tilde \rho_{g_0}=\tilde g$ is acceptable although there are no firm backgrounds 
to justify this very choice. 

The most informative is the relation (\ref{hldg1}): this relation not only suggests a necessity of subtraction
in the form factor $f$ but also determines the subtraction term. When considering the heavy--to--heavy
transitions we have found that the subtracted spectral density of the form (\ref{tildeffull})
\beq
\tilde f=\tilde f_D+(M_1^2-s_1+M_2^2-s_2)\tilde \rho_f
\eeq
matches the HQET expansion if $\tilde\rho_f\simeq \tilde g$ in LO. The eq. (\ref{hldg1}) prescribes 
$\tilde\rho_f=\tilde g$. 
\subsection{The scaling of the form factors with $m_Q$}
Let us point out that in the case of the heavy--to--light transition there is no parametrical suppression of the
anomalous contribution to the form factors compared with the normal one as it was in heavy--to--heavy transitions
and both parts contribute on equal footings. We shall present all the results for the normal part but the same
holds also for the anomalous part. 

In the region $\oq=O(1)$ one finds for the normal contribution
\begin{equation}
\label{scaling}
f_i(\oq)
=
\frac{1}{\sqrt{m_2}}
\int
\frac{ds_2\varphi(s_2)\lambda^{1/2}(s_2,m_1^2,m_3^2)}
{16\pi\,m_1}
\int\limits_{-1}^{1}
\frac
{d\eta \sqrt{\oq^2-1}\;\phi_0(z_1)\tilde f_i(z_1,z_2,m_1,m_2,m_3,\oq)}
{\sqrt{(m_1(\oq+1)+z_1+z_2+2m_3)(m_1(\oq-1)+z_1-z_2)}}
\end{equation}
where
\begin{eqnarray}
&&s_2=(m_1+m_3+z_2)^2,\nonumber\\
&&z_1=z_2\oq\left({1+\frac{2m_3+z_2}{m_1}}\right)+m_3(\oq-1)+\eta\sqrt{\oq^2-1}
\frac{\lambda^{1/2}(s_2,m_1^2,m_3^2)}{2m_1}+O(1/m_2). \nonumber
\end{eqnarray}
Using the relations (\ref{hlcoefficients}) one finds that 
the spectral densities scale at large $m_2$ as  
\begin{equation}
\tilde f_i=m_2^{n_i}\rho_i(\oq,m_1,m_3,z_1,z_2).
\end{equation}
Namely, 
\begin{eqnarray}
&
\tilde f_1=O(1),\quad 
\tilde f_2=O(m_2),\quad 
\tilde g=O(1),\quad 
\tilde a_1=O(1/m_2),\quad 
\tilde a_2=O(1),
\nonumber
\\
&
\tilde f=O(m_2),\quad 
\tilde g_1=O(1),\quad 
\tilde g_2=O(m_2),\quad 
\tilde g_0=O(1/m_2),\quad 
\tilde s=O(1).
\nonumber
\end{eqnarray}
Hence, the form factors have the scaling behaviour of the form
\begin{equation}
f_i=m_2^{n_i-1/2}r_i(\oq;m_1,m_3;\varphi_2,\phi_0).
\end{equation}
The variables $\oh$ and $\oq$ are connected with each other as follows
\begin{equation}
\oq=\frac{M_2}{m_1}\oh-\frac{\bla}{m_1}+O(1/m_2),
\end{equation}
and hence the ratio
\begin{equation}
f_i(\oh)/M^{n_i+1/2}
\end{equation}
is universal for the transition of any heavy meson into a fixed  light state. 
This behaviour reproduces the results of \cite{iwhl} up to the
logarithmic corrections which arise from the anomalous scaling of the quark currents in QCD. 

To summarize, matching the LO $1/m_2$ relations between the form factors in the quark model to 
the corresponding relations of \cite{iwhl} allowed us to determine the subtraction term in the form factor 
$f$ and equated to each other subtraction terms in the form factors $a_{1,2}$ and $g_0$. 
\newpage
\section{Numerical analysis}
In this section we apply the derived results to the model--dependent estimates of the universal form factors. 

\begin{figure}[hbt]
\begin{center}  
\mbox{\epsfig{file=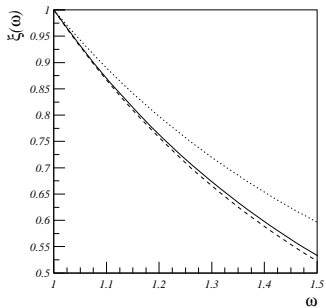,height=5.cm}  }
\end{center}
\caption{The Isgur--Wise function calculated with various quark--model parameters:
dashed line -- set 1, solid line -- set 2, dotted line -- set 3. 
\label{fig:iw}}
\end{figure}

1. First, let us study the dependence of the IW function on the parameters of the quark model. 
We adopt the exponential parametrization of 
the radial wave functions in the form $w(\vec k^2)\simeq \exp(-\vec k^2/2\beta^2)$. 
For obtaining the LO wave function $\phi_0$ we must take into
account the relationship between the parameters $z$ and $\vec k^2$ which is found from the following equation 
\beq
\sqrt{s}=\sqrt{\vec k^2+m_Q^2}+\sqrt{\vec k^2+m_3^2}=m_Q+m_3+z.
\eeq
This yields the relation 
\beq
\vec k^2=z(z+2m_3)+O(1/m_Q).
\eeq
Using the eq. (\ref{5vertex}) we come to the following form of the LO wave function 
\beq
\phi_0(z)\simeq\sqrt{\frac{z+m_3}{z+2m_3}}\exp\left(-\frac{z(z+2m_3)}{2\beta_0^2}\right),
\eeq
where $\beta_0$ is the LO harmonic oscillator parameter 
\beq
\beta(m_Q)=\beta_0+O(1/m_Q).
\eeq
For calculations we need to specify $\beta_0$. 
The exact value of $\beta_0$ is not known but extrapolating the known values of $\beta_D$, $\beta_B$, and
$\beta_{D^*}$ we find $\beta_0\simeq 0.4$ both in the WSB \cite{wsb} and the ISGW2 \cite{isgw2} models. 
In the model of ref. \cite{m1} from the analysis of the axial--vector decay constants of pseudoscalar 
mesons the following approximate dependence has been proposed
\beq
\beta(m_Q)\simeq 2.5\frac {m_Qm_3}{m_Q+m_3},
\eeq
which gives $\beta_0\simeq 2.5\; m_3$. Table \ref{table:parameters} presents the relevant numerical parameters. 
The results of calculating the IW function through the eq. (\ref{xi}) are shown in Fig. \ref{fig:iw}. 
Table \ref{table:parameters} demonstrates the values $\xi'(1)$ calculated with the eq. (\ref{rho2}) 
and the parameters of the quadratic fit of the form 
\beq
\xi(\oh)=1-\rho_i^2(\oh-1)+\delta_i(\oh-1)^2,
\eeq
obtained by interpolating the results of calculations in the range $1\le\oh\le1.5$. One can observe the value of $-\xi'(1)$ 
to be considerably larger than the parameter $\rho^2$ obtained by the interpolation
procedure. 
\begin{table}[hbt]
\caption{\label{table:parameters}
Parameters of the quadratic fit to the IW function and the NLO form factor $\xi_3$ in various quark--model versions.} 
\centering
\begin{tabular}{c||c|c||c|c|c||c|c|c|}
Ref.                & $m_3$ & $\beta_0$ & $-\xi'(1)$ & $\rho^2$ & $\delta$ & $\xi_3(1)$ & $\rho^2_{\xi_3}$ & $\delta_{\xi_3}$ \\
\hline\hline
Set 1 \cite{wsb}    & 0.35  &    0.4    &   1.47     &   1.34     &   0.78  &   0.077    &   1.49     &   0.88  \\
Set 2 \cite{isgw2}  & 0.33  &    0.4    &   1.43     &   1.31     &   0.76  &   0.08     &   1.4      &   0.84  \\
Set 3 \cite{m1}     & 0.25  &   0.63    &   1.37     &   1.12     &   0.62  &   0.17     &   1.2      &   0.65  
\end{tabular}
\end{table}

The IW function obtained from the dispersion quark model agrees with the results 
of the light--cone quark model applied to the elastic $P\to P$ transition for (infinitely) heavy mesons \cite{simula} 
$\xi(1.5)=0.62$ and turns out to be a bit smaller than the SR result $\xi(1.5)=0.66$ \cite{neubert93}.

The NLO form factor $\xi_3(\oh)$ is found to be very sensitive to the light--quark mass. 
The parameters of the quadratic fit to $\xi_3(\oh)$ calculated through eq. (\ref{xi3a}) in the form 
\beq
\label{fit}
h_i(\oh)=h_i(1)\;[1-\rho_i^2(\oh-1)+\delta_i(\oh-1)^2]
\eeq
for various sets of the quark--model parameters are shown in Table {\ref{table:parameters}. 
One can observe an approximate relation $\xi_3(\oh)/\xi(\oh)\simeq 0.08\;GeV$ for the light quark mass
$m_3\simeq0.35\;GeV$ and $\xi_3(\oh)/\xi(\oh)\simeq 0.17\;GeV$ for the light quark mass $m_3\simeq0.25\;GeV$ 
in the whole region $1\le\oh\le1.5$. 
This ratio is to be compared with the SR result with the $O(\alpha_s)$ corrections omitted \cite{neubert} 
\beq
\xi_3(\oh)/\xi(\oh)=\bla/3\simeq 0.16\; GeV,\quad \bla\simeq 0.5\;GeV. 
\eeq

2. We are in a position to estimate the higher order corrections as we can calculate the form factors at finite masses  
and the leading--order contribution separately. For the ISGW2 parameter set ($m_c=1.82$, $m_b=5.2$, 
$\beta_{D}=0.45$,  $\beta_{B}=0.43$, and $\beta_{D^*}=0.38$), the results on $h_{f_+}$ and $h_f$ 
are shown in Fig. \ref{fig:xi}, and Table \ref{table:xi} presents the results of interpolating in the range 
$1\le\oh\le1.5$ with a quadratic fit (\ref{fit}).
\begin{table}[hbt]
\caption{\label{table:xi}
The form factors calculated with the parameters of ISGW2 model.}
\centering
\begin{tabular}{|c||c|c|c|}
   & $h_{f_+}$ &$h_{f}$  & $\xi$ \\
\hline
\hline
$h(1)$   & 0.93  & 0.96 & 1.0  \\
$\rho^2$ & 0.87  & 1.06 & 1.25 \\
$\delta$ & 0.37  & 0.53 & 0.7  
\end{tabular}
\end{table}

\begin{figure}[hbt]
\begin{center}  
\mbox{\epsfig{file=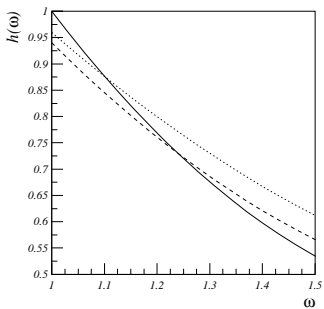,height=5.cm}  }
\end{center}
\caption{The velocity--dependent form factors calculated for the ISGW2 parameters:
solid line -- the IW function, dashed line -- $h_{f}(\oh)$, dotted
line -- $h_{f_+}(\oh)$. 
\label{fig:xi}}
\end{figure}
The form factor $h_f$ can be seen to agree with the results of a combined fit to 
the ALEPH, ARGUS, CLEO, DELPHI, and OPAL data \cite{jimack}
\beq
\rho^2_{f}=1.07\pm 0.20. 
\eeq
One observes sizeable difference between the form factors $h_f$ and $h_{f_+}$ and the Isgur--Wise 
function both in the value at zero recoil and the slope parameter. 
This indicates sizeable higher order corrections. At zero recoil the $1/m_Q$ corrections 
to $h_f$ vanish and writing \cite{neubert} 
\beq
\label{deltam2}
h_f(1)=1+\delta_{1/m^2}
\eeq
we find $\delta_{1/m^2}=-0.04$ that agrees well with the value $-0.055\pm0.025$ \cite{neubert}. 
At the nonzero recoil the higher--order corrections yield a more flat $\oh$--dependence of the form factor 
$h_f$ compared with the IW function such that 
the slope of the IW function in our model turns out to be considerably larger compared with the slope 
of $h_f$: An approximate relation $\rho^2_{IW}\simeq \rho^2_{h_f}-0.2$ is found. 
This is different from the SR result $\rho^2_{IW}\simeq \rho^2_{h_f}+0.2$ \cite{neubert}. 
Let us point out that $h'(1)$ turns out to be considerably 
larger than the parameter $\rho^2$ of the quadratic fit obtained by the interpolation over the range 
$1\le\oh\le1.5$ just as in the case of the Isgur--Wise function. }
\section{Conclusion}
We have performed a detailed analysis of the dispersion formulation of the quark model for meson decays  
based on representing the transition form factors as double spectral integrals in the 
invariant masses of the initial and final $q\bar q$ pairs through the soft wave functions of the 
initial and final mesons. The unsubtracted double spectral densities of the form factors at spacelike 
momentum transfers are calculated from the Feynman graphs, and subtractions in the spectral representations 
are determined by considering the two following cases: (i) by performing the heavy--quark expansion of the form factors in the case 
of heavy--to--heavy transitions and 
matching them to the HQET and (ii) by considering the relations between the form factors of the vector, axial--vector
and tensor currents in the case of the heavy--to--light transition and matching them to the general relations of 
ref. \cite{iwhl}. 
This procedure gives the spectral representations with appropriate subtractions at spacelike momentum 
transfers. 
The form factors at timelike momentum transfers are obtained by performing the analytical continuation in
$q^2$. The analytical continuation yields the appearance of the anomalous contribution to the form factors at $q^2>0$. 
As a result, we have come to an explicit model of the form factors describing the 
meson decays induced by the transition between finite--mass quarks which develops 
the correct structure of the heavy--quark expansion.  
Let us emphasize that the obtained representations allow us to calculate the form factors directly at the timelike region. 

Notice that for deriving the heavy--quark expansion we impose only a strong localization of the meson momentum
distribution with a width of order of the confinement scale. No other constraints on the soft wave functions 
or numerical parameters of the model have emerged. 

Our main results are:\\
1. We apply the method of ref. \cite{luke} to meson amplitudes of the tensor current and find the expansions of the
relevant form factors in LO and NLO in HQET. The $1/m_Q$ corrections in the form factor $h_{g_+}$ are found to vanish 
at zero recoil just as in $h_{f_+}$ and $h_f$. \\
2. The $1/m_Q$ expansion of the soft wave function is constructed. The wave function normalization condition 
which is a consequence of the vector current conservation in the full theory yields an infinite chain of normalization 
conditions for the wave function components emerging in various $1/m_Q$ orders. 
The LO normalization condition which is connected with the current conservation in the effective theory provides 
the correct normalization of the Isgur--Wise function at zero recoil. \\
3. In the LO all the transition form factors are represented through the Isgur--Wise function in accordance with HQET. 
The Isgur--Wise function is calculated through the LO component of the heavy meson wave function. 
The subtraction terms in spectral representations of the form factors do not contribute in LO. 
The anomalous contribution which emerges when the analytical continuation to the timelike region is performed comes into
the game only in the $1/m_Q^2$ vicinity of the of the zero recoil point but otherwise is suppressed by at least the second
inverse power of the heavy--quark mass. \\
4. The NLO analysis of the $P\to P$ transitions shows that the unsubtracted dispersion representations 
develop $1/m_Q$ structure in accordance with HQET and allows calculating $\xi_3$ through the LO wave function. \\
5. In the case of the $P\to V$ transition the unsubtracted representations for the form factors $g$, $g_1$, $g_2$, and 
$m_1a_2-m_2a_1$ have the NLO behavior compatible with HQET and yield the relation $\rho_1=\rho_2$ (or $\chi_3=0$) 
and $\chi_2=0$. At the same time, the unsubtracted form factors $f$, $g_0$, and $m_1a_2+m_2a_1$ do not agree with HQET 
in NLO. The matching procedure allows restricting the form of the subtraction terms. \\
6. Analyzing the heavy--to--light transitions and requiring the fulfillment of the Isgur--Wise relations
between the form factors of the tensor and vector and axial--vector currents \cite{iwhl} 
further constrains the subtraction terms. \\
7. We observe a discrepancy between the predictions of the various versions of the quark model on the NLO
universal form factors : Namely, the analysys of the WSB model \cite{wsb} performed in \cite{neubertr} resulted in $\rho_1\ne
\rho_2\ne 0$ 
and $\rho_3\ne 0$, but $\rho_4=0$; the quasipotential quark model \cite{faustov} predicts all the NLO form factors to be nonzero; 
a consistent relativistic 
treatment of only $q\bar q$ intermediate states in Feynman graphs performed in this paper gives 
$\rho_1=\rho_2$, $\rho_3$=0, and  $\rho_4\ne 0$. 
The origin of this discrepancy between the quark models is not fully understood and should be considered in
more detail. \\
8. The numerical results of the dispersion quark model using several parameter sets 
(i.e. constituent quark masses and wave functions) demonstrate a moderate dependence of the Isgur--Wise
functions on the light--quark mass and the shape of the wave function. 
Namely, the light--quark mass $m_3=0.25\div 0.35$ and the wave--function width $\beta_0=0.4\div 0.65$ 
yield $\xi(1.5)=0.55\div 0.6$ that is a bit smaller than the SR result $\xi(1.5)=0.66$ \cite{neubert93}. 

The form factor $h_f$ is found to have the behavior in agreement with predictions of other models and 
experimental results. We observe sizeable difference between the form factor $h_f$ and the Isgur--Wise 
function both in the value at zero recoil and the slope parameter. 

The size of the higher--order $1/m_Q$--corrections to $h_f(1)$, $\delta_{1/m^2}=-0.04$, 
obtained using the ISGW2 parameters agrees favorably with the sum rule estimates.

The slope of the IW function in our model turns out to be considerably larger compared with the slope 
of $h_f$: An approximate relation $\rho^2_{IW}\simeq \rho^2_{h_f}-0.2$ is found. 
This is opposite to the SR 
result $\rho^2_{IW}\simeq \rho^2_{h_f}+0.2$ \cite{neubert}. 
We would like to notice that the slope parameter
is very sensitive to the interpolation procedure: namely, the value $h'(1)$ turns out to be considerably 
larger than the result of the quadratic interpolation over the range $1\le\oh\le1.5$.

The dispersion quark model for the transition form factors can be further improved by
performing the heavy--quark expansion in higher orders and matching to the HQET order by order. 
It should be taken into account 
that the quark model effectively describes the whole $1/m_Q$ series, but the short--distance corrections 
are not contained in the model. The inclusion of such short--distance corrections into consideration 
is necessary for the application of the model to the analysis of the experimental results. 

\acknowledgements
I am grateful to I.Narodetskii, S.Simula, B.Stech and K.Ter--Martirosyan for discussions 
and to M.Neubert for helpful communication.
The work was supported by the RFBR under grants 95--02--04808a and 96--02--18121a.

\end{document}